\documentclass[amssymb,prb,twocolumn,showpacs]{revtex4}
\usepackage{amsmath,amsthm,amsfonts}  
\usepackage{epsf}
\usepackage{amssymb}  
\usepackage{graphicx}

\begin{document}
\title{Charge Transport Through Open, Driven Two-Level Systems with Dissipation}
\author{Tobias Brandes $^1$,  Ram\'on Aguado $^2$, and Gloria Platero $^2$}
\affiliation{1-Department of Physics, University of Manchester Institute of
Science and Technology (UMIST), P.O. Box 88, Manchester M60 1QD, United Kingdom}
\affiliation{2-Departamento de Teor\'{\i}a de la Materia Condensada,
Instituto de Ciencia de Materiales de Madrid,
CSIC, Cantoblanco 28049, Madrid, Spain.}
\date{\today{ }}

\begin{abstract}
We derive a Floquet-like formalism to calculate the stationary average current through an AC driven  
double quantum dot in presence of dissipation. The method allows us to take 
into account arbitrary coupling strengths both of a time-dependent field and a bosonic
environment. We numerical evaluate a truncation scheme and compare 
with analytical, perturbative results such as the Tien-Gordon formula.
\end{abstract}
\pacs{
73.23.Hk 
73.63.Kv 
03.65.Yz 
}
\maketitle
\section{Introduction}
Coupled quantum systems with small effective Hilbert spaces
are useful tools in order to study coherence, dissipation and the interaction
properties of  few-particle systems. In an electronic context,
an example are  coupled quantum dots \cite{Vaartetal95,Blietal96,Fujetal98,Blietal98b,Taretal99}, where 
strong  interactions between electrons  \cite{LD98,BLD99,BL00} 
define a Coulomb blockade regime with tunnel-splitted many-body ground
states separated from the
remaining excited states. The ultimate limit of two states defines
a two-level system for the charge degree of freedom, with electrons of a fixed spin 
tunneling between two quantum dots. Studying transport and dissipation 
then leads to a non-equilibrium or `open' (pseudo)spin-boson problem,
where the coupling to external reservoirs opens the path to investigate
properties such as shot noise \cite{AB03}
or decoherence in a controllable semiconductor environment.

Additional insight into the quantum dynamics of electrons can be gained by 
making the parameters of the problem time-dependent. When the time-dependence
is slow, this can give rise to 
a variety of adiabatic phenomena such as charge pumping 
\cite{Geretal90,Kouetal91,Potetal92,Grabert,Swietal99,Bro98,PB01,CB02,MB01,RB01},
adiabatic control of state vectors \cite{Bonetal98,BRB01},
or operations relevant for quantum information processing in a condensed-matter
setting \cite{Ave98,Ave99,NPT99,SLM01,MSS01,EL01,TH02}. 
Different physics occurs in the  high frequency regime where
 monochromatic time-variation  
induces photo-excitations, such as for
coupling of AC fields to quantum dots
\cite{BS94,HS95,SW96,BBS97,SN96,SWL00},
which has been tested experimentally \cite{Kouetal94,Oosetal98,Blietal98a,Holetal00,Qinetal01,QHEB01} recently.

In general, AC driven systems \cite{Shi65,hen68} and their application to various 
mesoscopic transport 
\cite{KM89,BWK93,FD93,BPT93,HH96,MG96,Bue00,Zhang02a}
and tunneling \cite{TG63,Groetal91,JW92,IPT94,JWM94,Ignetal95,Wag95,AIP96,Wag96,WZ97,WS99,Lehetal02,Cametal03}
regimes
have quite a long history, although
the inclusion of 
interactions and correlations is a relatively new area. 
In low-dimensional systems,
investigations have concentrated on  one-dimensional models \cite{Sk96,CSK98,Fecetal01,Pham03,Vicetal02}, 
the modification of Kondo-resonances by
AC fields \cite{LAPT98,KNG99}, mean-field type approximations \cite{AP98},
or exact studies of driven few-electron systems \cite{CP02,CP02a}.

In this paper, we combine AC driving with the {\em dissipative dynamics} of a 
two-level system (double quantum dot) under transport conditions, i.e. in a situation
where electrons in the Coulomb blockade regime 
can tunnel from reservoirs into and off two tunnel-coupled quantum dots,
with the possibility to absorb from or emit bosons into a heat bath while simultaneously
interacting with a classical time-periodic electrical field. 
At first sight, combining such a multitude of possible interactions within one and the same
model might look unsuitable for a useful theoretical discussion. However, as we will 
demonstrate in this paper, it is possible 
to 
calculate experimentally relevant observables such as the time-averaged stationary
current, with the help of the (heat bath) boson spectral density 
$J(\omega)$ as single, main input of the theory only. In particular, we show how 
within the polaron transformation approach and for a 
given $J(\omega)$, one can calculate the current for arbitrarily 
strong coupling to bosonic modes {\em and} an AC field. 

The paper is organized as follows: in section II, we describe the model Hamiltonian and 
derive a Floquet-like formalism for the stationary density operator. In section III, we 
compare analytical results for limiting cases with numerical data, and conclude with a short 
discussion and an outlook in section IV.

\section{Master Equation Formalism for AC Driven Double Quantum Dots}
In the following, we shall develop the general framework leading to 
explicit expressions for the stationary current through dissipative, driven double quantum dots.
Our approach is in part similar to the treatment of closed, dissipative two-level systems with AC
driving as reviewed by Grifoni and H\"anggi \cite{GH98}. Here, we generalise this 
approach to take into account tunneling between the dots and the leads. In the non-dissipative
case, this problem was treated by Gurvitz and Prager \cite{GP96,Gur98} for non-driven double dots, and 
for coherently AC driven double dots by Stoof and Nazarov \cite{SN96}.

\subsection{Model Hamiltonian}
We assume that the driven two-level system is
defined in a double quantum dot device \cite{Fujetal98}.
In the regime of strong Coulomb
blockade, these
can be tuned into a regime where the internal dynamics is governed by a time-dependent (pseudo)
spin-boson  model (dissipative two-level system
\cite{Legetal87}), ${\cal H}_{SB}(t)$.
The latter 
describes one additional `transport' electron which
tunnels between a left (L) and a right (R) dot with {\em time-dependent} energy
difference $\varepsilon(t)$ and inter-dot coupling $T_c(t)$, and is
coupled to a dissipative bosonic bath (${\cal H}_B= \sum_{\bf
Q}\omega_{Q} a^{\dagger}_{\bf Q} a_{\bf Q}$),
\begin{eqnarray}\label{modelhamiltonian}
{\cal H}_{SB}(t)&=& \Big[\frac{\varepsilon(t)}{2}
+\sum_{\bf Q} \frac{g_{Q}}{2} \left(a_{-\bf Q} + a^{\dagger}_{\bf
Q}\right)\Big]
\hat{\sigma}_z\nonumber\\
&+& T_c(t)
\hat{\sigma}_x
+{\cal H}_B.
\end{eqnarray}
The effective Hilbert space of double dot (without any coupling to electron leads
or bosons) then consists of two 
(many-body) states $|L \rangle=|N_L+1,N_R \rangle$ and $|R \rangle=|N_L,N_R+1
\rangle$ and is defined by a pseudospin
$\hat{\sigma}_z \equiv |L \rangle \langle L|-|R \rangle \langle
R|\equiv \hat{n}_L-\hat{n}_R$ and $\hat{\sigma}_x\equiv |L \rangle \langle
R|+|R \rangle \langle L|\equiv \hat{p}+\hat{p}^{\dagger}$. 

The effects
of the bosonic bath are fully described as usual by  a spectral density
\begin{eqnarray}\label{Jdef}
J({\omega})\equiv\sum_{\bf Q} |g_{Q}|^2\delta(\omega-\omega_Q),  
\end{eqnarray}
where $\omega_Q$ are the frequencies of the bosons and the $g_Q$
denote interaction constants.
When showing particular results we will be using
\begin{eqnarray}
J(\omega)={2\alpha}{\omega} e^{-\omega/\omega_c},   
\end{eqnarray}
corresponding 
to a generic Ohmic bath. More realistic forms can be easily incorporated into our formalism but
for simplicity, in this work we restrict ourselves to the Ohmic case.

The coupling to
external free electron reservoirs ${\cal H}_{res}=\sum_{k_\alpha}\epsilon_{k_\alpha}
c_{k_\alpha}^{\dagger}c_{k_\alpha}$
is described by the usual tunnel Hamiltonian
\begin{eqnarray}
  {\cal H}_T=\sum_{k_\alpha{}}
(V_k^\alpha{} c_{k_\alpha{}}^{\dagger}s_\alpha{}+H.c.), 
\end{eqnarray}
$\hat{s}_\alpha=|0 \rangle
\langle \alpha|$ ($\alpha$=L,R). Here, a third
state $|0 \rangle=|N_L,N_R \rangle$ 
describes an `empty' DQD. Its presence leads
to strong modifications both in the mathematical description
as well in the physics of this problem, as compared to the 
case of an isolated spin-boson Hamiltonian.
Here, the reservoir-related parameters of ${\cal H}_{res}$ and  ${\cal H}_T$ have been assumed 
to be time-independent which again is an approximation which might not be always fulfilled in
experiments. Again, we concentrate on the simplest possible case in this work and neglect the 
effect of, e.g., a time-dependence in the external electro-chemical potentials.

The full model as described by 
\begin{eqnarray}
  {\cal H}(t)= {\cal H}_{SB}(t)+{\cal H}_{res}+{\cal H}_{T}
\end{eqnarray}
now offers the possibility to study
non-equilibrium properties of a time-dependent, `open' dissipative two-level
system. 
Note that in spite of the third, `empty' state
$|0\rangle$ we continue to use the term `two-level system' here and in the following: although
the presence of $|0\rangle$ leads to strong modifications of, e.g.,  the equations
of motion of the density operator, it turns out that
the internal dynamics of the system is still closely related to that of the dissipative spin boson
problem.

The time-dependent spin-boson problem is in general characterised
by the fact that  {\em both} $\varepsilon(t)$ and $T_c(t)$ are time-dependent.
One  can  then
investigate interesting effects such as adiabatic charge pumping, dissipative
Landau-Zener tunneling \cite{BV02}, or for the closed system (no coupling to the leads)
the control of quantum superpositions \cite{Hayetal03}. Although this general time-dependence offers
the richest spectrum of possible physical phenomena, one is clearly 
strongly restricted by the fact that nearly no analytical solutions
are available.
In this paper, our goal is to develop a systematic theory for 
the stationary state of a somewhat simpler situation, i.e., the case
where $T_c(t)\equiv T_c$ is constant, with the time-dependence solely 
contained in the bias  $\varepsilon(t)$.

\subsection{Equations of Motion}
In the following, we treat the coupling to the reservoirs 
within the Born and Markov approximation with respect to ${\cal H}_{T}$
\cite{BK99,SN96}, such that higher order effects like
cotunneling or the Kondo effect are not considered. This Born-Markov
approximation becomes exact in the limiting case of infinite source-drain voltage \cite{GP96}.
Specifically, one sets
the Fermi distributions for the left (right) reservoir $f_L=1$ ($f_R=0$) whence the chemical potentials
of the leads no longer play any role.
Furthermore, the tunnel rates which are given by
\begin{eqnarray}
\Gamma_{\alpha}=2\pi\sum_{k_{\alpha}}|V_k^{\alpha}|^2\delta(\epsilon-\epsilon_{k_{\alpha}}),
\quad \alpha=L/R,
\end{eqnarray}
are assumed to be independent of energy.
We mention that the generalisation to intermediate voltage regimes (finite bias) for double dots is 
a difficult and non-trivial problem even in the undriven case, which is why we only discuss
the infinite-bias limit in this paper.

The derivation of the equations of motion for the dot observables is now
very similar to the non-driven case \cite{BK99}. 
The time-dependence of the Hamiltonian enters via
the replacement of the  phase factors 
$ e^{i\varepsilon(t-t')}$ in the free un-driven 
time-evolution of the dots, by $e^{i\int_{t'}^{t}ds\,\varepsilon(s)}$ for the driven case.
Introducing the vectors ${\bf A}\equiv
(\hat{n}_L,\hat{n}_R,\hat{p},\hat{p}^{\dagger})$, ${\bf
\Gamma}=\Gamma_L{\bf e}_1$ (${\bf e}_1,...,{\bf e}_4$ are unit vectors)
and a time-dependent matrix memory kernel ${M}$,
the equations of motion (EOM) can be formally written as \cite{AB03}
[$\langle..\rangle\equiv {\rm Tr} ..\rho(t)]$,
\begin{equation}\label{expectation}
  \langle {\bf A}(t)\rangle   \!=\! \langle {\bf A}(0)\rangle
 + \int_{0}^{t}dt' \left\{{M}(t,t')\langle {\bf A}(t')\rangle  + {\bf \Gamma} \right\}.
\end{equation}
This formulation is a useful starting point for, e.g., the calculation of shot noise. Note that
in contrast to the undriven case, the memory kernel $M$ depends on both times $t$ and $t'$ because
there is no time-translation invariance in presence of driving. Explicitely, 
the equations for the dot expectation values read 
\begin{widetext}
\begin{eqnarray}\label{eom3new}
\frac{\partial}{\partial t}\langle n_L\rangle_t&=&-iT_c\left\{
\langle p \rangle_{t}-\langle p^{\dagger}\rangle_{t} \right\} 
+\Gamma_L\left[1-\langle n_L\rangle_{t} - \langle n_R\rangle_{t}\right]
\nonumber\\
\frac{\partial}{\partial t}\langle
n_R\rangle_t&=&iT_c \left\{ \langle
p\rangle_{t}-\langle p^{\dagger}\rangle_{t} \right\} 
-{\Gamma}_R\langle n_R\rangle_{t} 
\nonumber\\ 
\langle p\rangle_t&=& -\int_0^tdt'   e^{i\int_{t'}^{t}ds\,\varepsilon(s)}
\left[ \left( \frac{\Gamma_R}{2}\langle {p}\rangle_{t'}
+ iT_c  \langle n_L\rangle_{t'} \right) C(t-t')- iT_c \langle
n_R\rangle_{t'} C^*(t-t') \right] \nonumber\\
\langle p^{\dagger}\rangle_t&=& -\int_0^tdt'   e^{-i\int_{t'}^{t}ds\,\varepsilon(s)}
\left[ \left( \frac{\Gamma_R}{2} \langle {p}^{\dagger}\rangle_{t'}
- iT_c  \langle n_L\rangle_{t'} \right) C^*(t-t')+ iT_c \langle
n_R\rangle_{t'} C(t-t') \right].
\end{eqnarray}
\end{widetext}
Here, half the decay rate (tunnel rate $\Gamma_R/2$) of the system
appears in the off-diagonal terms $p$ and $p^{\dagger}$, acting as a source of dephasing due to
tunneling of an electron {\em out of} the dot in either direction.
Furthermore, the boson correlation function for a harmonic bath with spectral density $J(\omega)$,
Eq.~(\ref{Jdef}) and at equilibrium temperature $k_BT=1/\beta$ enters,
\begin{eqnarray}
  C(t)&\equiv&e^{-Q(t)}\\
Q(t)\!&\equiv&\!\int_0^{\infty}d\omega
\frac{J(\omega)}{\omega^2} \left[ \left(1- \cos \omega t\right)
\coth \left(\frac{\beta \omega}{2}\right) + i \sin \omega t
\right]\nonumber.
\end{eqnarray}
In deriving, the equations for the off-diagonal elements $\langle p^{(\dagger)}\rangle$, 
we used the polaron transformation (POL) and factorised the bosonic correlation
functions from the dot operators in the equations of motions for the reduced 
density operator of the (pseudo) spin-boson system. This means that \ Eq.~(\ref{eom3new}) is
perturbative (though to infinite order) in the interdot coupling $T_c$. 

Alternatively, one can perform a perturbation theory in the electron-boson coupling $g_Q$, (weak coupling  
`PER' approach). In a calculation for
an undriven double quantum dot, both approaches have been compared recently for the
stationary current \cite{BV01} and the frequency-dependent current noise \cite{AB03}.
For the spin-boson problem with $\Gamma_{R/L}=0$, 
it is well-known that POL is equivalent to a double-path integral `non-interacting blip approximation' 
(NIBA) that works
well for zero bias $\varepsilon=0$ but for $\varepsilon\ne 0$  does not coincide with PER at small couplings 
and very low temperatures. 
PER works in the correct  bonding and anti-bonding eigenstate basis of the  hybridized system,
whereas the energy scale $\varepsilon$ in POL is that of the two isolated dots ($T_c=0$). 
This difference reflects the general dilemma of two-level-boson Hamiltonians: either one is 
in the correct basis of the hybridized two-level system and perturbative in $g_Q$, or
one starts from the `shifted oscillator' polaron picture that becomes correct for
$T_c=0$. In fact, the polaron (NIBA) approach does not coincide with
standard damping theory \cite{Weiss} because it does not
incorporate the  square-root, non-perturbative in $T_c$
hybridization form of the level splitting $\Delta= \sqrt{\varepsilon^2+4T_c^2}$. 
However, for large $|\varepsilon| \gg T_c$,
$\Delta \to |\varepsilon|$, and POL and PER turn out to agree very well for the undriven case \cite{BV01}.

\subsection{Stationary Quantities}
In a quantum system that is continuously driven by an external, time-dependent source, 
stationary quantities can be defined for expectation values approaching
a fixed point or a quasi-stationary, periodic motion
for large times $t$. In particular, we will be interested in quantities like the time-averaged
electronic current. It is then useful to 
split the time-dependent part off $\varepsilon(t)$ as
\begin{eqnarray}
  \label{eq:epsdef1}
  \varepsilon(t)=\varepsilon + \tilde{\varepsilon}(t),
\end{eqnarray}
and to introduce the
Laplace transform $\hat{f}(z)=\int_{0}^{\infty}dt e^{-zt}f(t)$ of a function $f(t)$.
The time-evolution of the isolated spin-boson system for $T_c=0$ is governed by
the correlation function $C(t)=C^*(-t)$. The Laplace transform of these,
\begin{eqnarray}
  \label{eq:Laplacedef}
  \hat{C}_{\varepsilon}(z)&\equiv&\int_{0}^{\infty}dt e^{-zt}e^{i\varepsilon t}C(t)\nonumber\\
  \hat{C}_{\varepsilon}^*(z)&\equiv&\int_{0}^{\infty}dt e^{-zt}e^{-i\varepsilon t}C^*(t).
\end{eqnarray}
defines  free propagators for quasiparticles  in the uncoupled dots and in absence of coupling 
to electron reservoirs. In absence of electron-boson coupling, this simply describes the free 
time evolution of a particle described by the diagonal Hamiltonian $\varepsilon \hat{\sigma}_z$, 
whereas for non-zero boson coupling these become `dressed' polarons.
In addition, the decay via the right reservoir at rate $\Gamma_R$ leads to
a finite quasiparticle life-time and consequently
a renormalisation of the propagators as
\begin{eqnarray}
  \label{eq:Ddef}
 \hat{D}_{\varepsilon}(z)&\equiv&\frac{\hat{C}_{\varepsilon}(z)}{1+\Gamma_R\hat{C}_{\varepsilon}(z)/2},\quad
 \hat{E}_{\varepsilon}(z)\equiv\frac{\hat{C}^*_{-\varepsilon}(z)}{1+\Gamma_R\hat{C}_{\varepsilon}(z)/2}\nonumber\\
 \hat{D}^*_{\varepsilon}(z)&\equiv&\frac{\hat{C}^*_{\varepsilon}(z)}{1+\Gamma_R\hat{C}^*_{\varepsilon}(z)/2},\quad
 \hat{E}^*_{\varepsilon}(z)\equiv\frac{\hat{C}_{-\varepsilon}(z)}{1+\Gamma_R\hat{C}^*_{\varepsilon}(z)/2},\nonumber\\
\end{eqnarray}
These expressions appear in the 
calculation in Appendix A, where
Eq.~(\ref{eom3new}) is solved for the coherences $\langle p\rangle$ and 
$\langle p^{\dagger}\rangle$
in order to obtain two closed equations for the occupancies $\langle n_{L/R} \rangle$,
\begin{widetext}
\begin{eqnarray}
  \label{eq:eomLapl3}
  z\hat{n}_L(z)-\langle n_L \rangle_0 &=& - \int_{0}^{\infty}dt e^{-zt}
\left[ \langle n_L\rangle_{t} \hat{K}(z,t)-\langle n_R\rangle_{t} \hat{G}(z,t)\right]
+\Gamma_L\left[ \frac{1}{z} - \hat{n}_L(z)- \hat{n}_R(z) \right]
\nonumber\\
z\hat{n}_R(z)-\langle n_R \rangle_0 &=&  \int_{0}^{\infty}dt e^{-zt}
\left[ \langle n_L\rangle_{t} \hat{K}(z,t)-\langle n_R\rangle_{t} \hat{G}(z,t)\right]
-{\Gamma}_R \hat{n}_R(z)
\nonumber\\
\hat{K}(z,t)&\equiv&\int_{0}^{\infty}dt'e^{-zt'}
\left[T_c(t+t')T_c^*(t){D}_{\varepsilon}(t')+
T_c^*(t+t')T_c(t){D}^*_{\varepsilon}(t')\right]\nonumber\\
\hat{G}(z,t)&\equiv&\int_{0}^{\infty}dt'e^{-zt'}
\left[T_c(t+t')T_c^*(t){E}_{\varepsilon}(t')+
T_c^*(t+t')T_c(t){E}^*_{\varepsilon}(t')\right],
\end{eqnarray}
\end{widetext}
where here and in the following we omit the $\langle.. \rangle$ in the Laplace 
transformed expectation values to simplify the notation, and 
we defined
\begin{eqnarray}
  T_c(t)&\equiv&T_ce^{+i\int_{0}^{t}ds\,\tilde{\varepsilon}(s)},\quad
  T_c^*(t)\equiv T_ce^{-i\int_{0}^{t}ds\,\tilde{\varepsilon}(s)}.
\end{eqnarray}
Up to here the transformations have been valid for an arbitrary time-dependence in $\varepsilon(t)$.
>From now  on, we specify to the time-periodic form
\begin{eqnarray}
  \varepsilon(t)=\varepsilon(t+2\pi/\Omega),
\end{eqnarray}
where $2\pi/ \Omega$ is the period of the time-dependent field (we further specify 
to a sinusoidal time-dependence of $\varepsilon(t)$ below).

We expect the system to approach an asymptotic
quasi-stationary state. Then, the time-evolution
of all quantities $f(t)$ can be decomposed into Fourier series
\begin{eqnarray}
  \label{eq:stat1}
  f(t)\to f^{as}(t)=\sum_ne^{-in\Omega t}f_n,
\end{eqnarray}
with multiples of the angular frequency $\Omega$ 
of the external field. 
Following Grifoni and H\"anggi \cite{GH98}, we decompose
$\hat{K}(z,t)$ and $\hat{G}(z,t)$
into Fourier series,
\begin{eqnarray}\label{eq:KGdec}
\hat{K}(z,t)&=&\sum_m {K}_m(z)e^{-i m \Omega t}\nonumber\\
\hat{G}(z,t)&=&\sum_m {G}_m(z)e^{-i m \Omega t}.
\end{eqnarray}
The  corresponding Fourier expansions
$\langle n_L \rangle^{asy}_t\equiv\sum_m \nu_me^{-im\Omega t}$ and
$\langle n_R \rangle^{asy}_t\equiv\sum_m \mu_me^{-im\Omega t}$ of 
the {\em asymptotic} occupancies
can then easily be Laplace transformed,
\begin{eqnarray}
\hat{n}_L^{asy}(z)=\sum_m \frac{\nu_m}{z+im\Omega},\quad
\hat{n}_R^{asy}(z)=\sum_m \frac{\mu_m}{z+im\Omega}
\end{eqnarray}
and inserted back into  Eq.~(\ref{eq:eomLapl3}). Comparing the complex poles at $z=-iM\Omega$
in the two equations for $\hat{n}_L(z)$ and $\hat{n}_R(z)$ and assuming
that  $K_m(z)$ and $G_m(z)$ are regular there, one obtains
an infinite system of linear equations for the Fourier coefficients $\nu_m$ and $\mu_m$,
\begin{widetext}
\begin{eqnarray}
  \label{eq:fourier2}
 -iM\Omega
{\nu_M}&=&-\sum_{n}\left[\nu_nK_{M-n}(-iM\Omega)-\mu_n G_{M-n}(-iM\Omega)\right]
+\Gamma_L\left[\delta_{M,0}-\nu_M-\mu_M\right]\nonumber\\
\left[{\Gamma}_R -iM\Omega\right]{\mu_M}&=&\phantom{-}\sum_{n}\left[\nu_n K_{M-n}(-iM\Omega)-\mu_n G_{M-n}(-iM\Omega)\right].
\end{eqnarray}
\end{widetext}
Upon adding these two equations, one has
\begin{eqnarray}
   \label{eq:lowesttc2}
 -\frac{\mu_M}{\nu_M} \equiv r_M \equiv \left[1
+\frac{\Gamma_R}{\Gamma_L-iM\Omega}\right]^{-1}, \quad M\ne 0,
\end{eqnarray}
and Eq. (\ref{eq:fourier2}) can be transformed into a single matrix equation for the coefficients $\nu_n$,
\begin{eqnarray}\label{lineardef}
  \sum_{n=-\infty}^{\infty} A_{mn}\nu_n = b_m,
\end{eqnarray}
where
\begin{eqnarray}\label{Aequation}
  A_{mn} &\equiv& \left( \Gamma_L -in\Omega -r_n\Gamma_L\right)\delta_{m,n}\nonumber\\
&+& K_{m-n}(-im\Omega) + r_nG_{m-n}(-im\Omega)\nonumber\\
b_m&\equiv& \frac{\Gamma_L{\Gamma}_R}{\Gamma_R + \Gamma_L}\delta_{m0} + 
\frac{\Gamma_L}{\Gamma_R + \Gamma_L}G_{m}(-im\Omega).
\end{eqnarray}

\subsection{Charge Current}
In the Master equation approach, 
the expectation values of the electron current through the double dot is obtained 
in a fairly easy manner. One has to consider the average charge 
flowing through one of the three intersections, i.e.,  left lead/left dot, left dot/right dot, 
and right dot/right lead. This gives rise to the three corresponding electron currents
$I_L(t)$, $I_R(t)$, and the interdot current $I_{LR}(t)$. 
>From the equations of motion, \ Eq.~(\ref{eom3new}), one recognises that 
the temporal change of the occupancies $\langle n_{L/R}\rangle_t$ is due
to the sum of an `interdot' current $\propto T_c$ and a `lead-tunneling' part.
Specifically, the current from left to right through the left (right) tunnel barrier is
\begin{eqnarray}\label{ILandR}
  I_L(t) &=& -e \Gamma_L\langle n_0\rangle_t = 
-e\Gamma_L\left[1-\langle n_L\rangle_{t} - \langle n_R\rangle_{t}\right]
\nonumber\\
I_R(t) &=& -e \Gamma_R \langle n_R\rangle_{t},
\end{eqnarray}
and the interdot current is
\begin{eqnarray}\label{ILR}
  I_{LR}(t) &=& -ieT_c\left\{
\langle p \rangle_{t}-\langle p^{\dagger}\rangle_{t} \right\}\\
&=& -e \frac{\partial}{\partial t}\langle n_R\rangle_t +I_R(t)
= e \frac{\partial}{\partial t}\langle n_L\rangle_t  +I_L(t)\nonumber.
\end{eqnarray}
In the stationary case, all the three currents are the same: adding the two equations
Eq.~(\ref{eq:fourier2}) for $M=0$, we first obtain
\begin{eqnarray}
  \Gamma_R\mu_0 = \Gamma_L(1-\mu_0-\nu_0).
\end{eqnarray}
Using furthermore the Fourier expansion of $\langle n_{L/R} \rangle^{asy}_t$, we recognize from
Eq.~(\ref{ILandR},\ref{ILR}) that
\begin{eqnarray}
  \bar{I} = \overline{I_L(t)} = \overline{I_{LR}(t)} = \overline{I_R(t)} = -e\Gamma_R \mu_0,
\end{eqnarray}
where the bar denotes the temporal average of the asymptotic quantities over 
one period $\tau\equiv 2\pi/\Omega$. This simple result means that the 
stationary current is determined by the Fourier component $\mu_0$ only. Note, however, that $\mu_0$ 
is part of the solution of an infinite set of the linear equations  Eq.~(\ref{lineardef}). 
Using  Eq.~(\ref{eq:fourier2}) and  Eq.~(\ref{eq:lowesttc2}), one can express $\mu_0$ in terms of $\mu_{N\ne0}$
for the alternative expression 
\begin{eqnarray}\label{currentstationary}
\bar{I}=-e\Gamma_R
  \frac{K_0(0)-\sum_{n\ne 0}\left[
K_{-n}(0)/r_n+ G_{-n}(0)\right]\mu_n}
{\Gamma_R+K_0(0)/r_0+G_0(0)}.
\end{eqnarray}
This form is in particular useful for the discussion of the Tien-Gordon limit below.

\subsection{Sinusoidal Time-Dependence}
In the following, we specify the time-dependence of the bias
$\varepsilon(t)$ to a monochromatic sinusoidal field
\begin{eqnarray}
  \label{eq:eeesin}
  {\varepsilon}(t)=\varepsilon + \Delta \sin(\Omega t),
\end{eqnarray}
where the constant part is denoted by $\varepsilon$. 
We introduce the notation,
\begin{eqnarray}\label{Cepsdefine}
  \hat{C}_{\varepsilon}(-i\omega)&=&
\hat{C}_{\varepsilon+\omega}(z=0^+)\equiv \hat{C}_{\varepsilon+\omega}\nonumber\\
 \hat{C}^*_{\varepsilon}(-i\omega)&=&
\hat{C}^*_{\varepsilon-\omega}(z=0^+)\equiv \hat{C}^*_{\varepsilon-\omega}
\end{eqnarray}
and correspondingly for the propagators $D$, $D^*$, $E$, and $E^*$,  Eq.(\ref{eq:Ddef}).
Then, invoking the decomposition of the phase factor into Bessel functions,
\begin{eqnarray}
    \label{eq:expdecomp1}
& & T_c(t+t')T_c^*(t') = T_c^2e^{i\int_t^{t+t'}ds {\Delta \sin(\Omega s)}}=\\
& &T_c^2\sum_{nn'}i^{n'-n}J_n\left(\frac{\Delta}{\Omega}\right)J_{n'}\left(\frac{\Delta}{\Omega}\right)
e^{-in\Omega t'} e^{-i(n-n')\Omega t}\nonumber
\end{eqnarray}
and the definitions of the Fourier components
${K}_m(z)$ and ${G}_m(z)$, cf. Eq. (\ref{eq:KGdec}) and Eq. (\ref{eq:eomLapl3}), one obtains
\begin{widetext}
\begin{eqnarray}\label{KGexplicit}
  {K}_m(-im'\Omega)&=& i^{-m}T_c^2\sum_n
\left[ J_n\left(\frac{\Delta}{\Omega}\right)J_{n-m}\left(\frac{\Delta}{\Omega}\right)
\hat{D}_{\varepsilon+(m'-n)\Omega}
+J_n\left(\frac{\Delta}{\Omega}\right)J_{n+m}\left(\frac{\Delta}{\Omega}\right)
\hat{D}^*_{\varepsilon-(m'+n)\Omega}\right]\nonumber\\
 {G}_m(-im'\Omega)&=& i^{-m}T_c^2\sum_n
\left[ J_n\left(\frac{\Delta}{\Omega}\right)J_{n-m}\left(\frac{\Delta}{\Omega}\right)
\hat{E}_{\varepsilon+(m'-n)\Omega}
+J_n\left(\frac{\Delta}{\Omega}\right)J_{n+m}\left(\frac{\Delta}{\Omega}\right)
\hat{E}^*_{\varepsilon-(m'+n)\Omega}\right].
\end{eqnarray}
\end{widetext}

\section{Analytical Results}
In the following, we first discuss the limits where analytical results for the stationary current
$\bar{I}$ can be obtained, and then turn to a comparison with numerical calculations.

\subsection{Time-Independent Case}
For $\Delta=0$, i.e. in absence of the time-dependent (driving) part in $\varepsilon(t)$, 
we recover  previous results \cite{BK99} for stationary transport in 
dissipative double quantum dots. One then has 
$\hat{K}(z,t)=\hat{K}(z)$ and $\hat{G}(z,t)=\hat{G}(z)$ such that
$  K_n(z)=G_n(z)=0$ for $n\ne 0$.
Using
$ K_0(0)=2{\rm Re}[T_c^2\hat{C}_{\varepsilon}/({1+\Gamma_R \hat{C}_{\varepsilon}/2})]$,
together with
$  G_0(0)=2{\rm Re}T_c^2[{\hat{C}^*_{-\varepsilon}}/({1+\Gamma_R \hat{C}_{\varepsilon}/2})]$,
%
after some algebra we re-derive the  previous result \cite{BK99}  for the stationary current,
\begin{eqnarray}\label{currentstat}
& &\overline{I} =-eT_c^2\frac{2\mbox{\rm Re}(\hat{C}_{\varepsilon})+\Gamma_R|\hat{C}_{\varepsilon}|^2}
{|1+\Gamma_R\hat{C}_{\varepsilon}/2|^2+2T_c^2B_{\varepsilon}}\\
B_{\varepsilon}&\equiv&
\mbox{\rm Re}\left\{(1+\Gamma_R\hat{C}_{\varepsilon}/2)\left[
 \frac{\hat{C}_{-\varepsilon}}{\Gamma_R}+\frac{\hat{C}_{\varepsilon}^*}
{\Gamma_L}\left(1+\frac{\Gamma_L}{\Gamma_R}\right)\right]\right\}\nonumber
\end{eqnarray}
(note the absence of the factor $2$ in the definition of the rates here \cite{BK99}).
The result  Eq.~(\ref{currentstat}), which can be compared \cite{BV01} to an 
alternative derivation using perturbation theory in the boson coupling
$\alpha$, 
generalizes the case of elastic tunneling through double quantum dots to
inelastic tunneling with coupling to an arbitrary bosonic heat bath.
For $\alpha=0$, we re-derive the Stoof-Nazarov expression for the
stationary current without dissipation,
\begin{eqnarray}\label{SNcurrent}
  \overline{I}_{\alpha=0} =-e\frac{T_c^2\Gamma_R}{\varepsilon^2+\Gamma_R^2/4+T_c^2(2+\Gamma_R/\Gamma_L)}.
\end{eqnarray}

\subsection{Lowest order $T_c^2$: Tien-Gordon Result}
In the time-dependent case, we are able to derive analytical results by considering
the limit of small interdot coupling $T_c$, or large frequencies $\Omega$. These
two limits do not yield identical results because apart from $T_c$ and $\Omega$, 
there are four other  energy scales (bias $\varepsilon$, rates $\Gamma_{L}$, $\Gamma_{L}$,
boson cut-off $\omega_c$) in the problem. 

Considering \ Eq.~(\ref{SNcurrent}) for the undriven, non-dissipative current,
lowest order perturbation theory in $T_c$ is valid for $ T_c\sqrt{2+\Gamma_R/\Gamma_L}\ll
\Gamma_R,|\varepsilon|$. The additional energy scale $\Omega$ due to
AC-driving requires that this condition is generalised to
\begin{eqnarray}
  T_c\sqrt{2+\frac{\Gamma_R}{\Gamma_L}}\ll \Omega,\Gamma_R,|\varepsilon+n\Omega|,\quad n=\pm 0,1,2,..,
\end{eqnarray}
which indicates that at the resonance points $\varepsilon=n\Omega$ such a  perturbation theory must break down,
as is corroborated by our numerical results discussed below.

Considering the expression for $\mu_M$ in   Eq.~(\ref{eq:fourier2}), one recognises that
$\mu_M=O(T_c^2)$ because the Fourier components of the functions $K$ and $G$ are proportional to
$T_c^2$, cf.  Eq.~(\ref{eq:eomLapl3}). Owing to the full expression 
Eq.~(\ref{currentstationary}), the stationary current 
in lowest order of $T_c$ is $\overline{I}=\overline{I}^{\rm TG}+O(T_c^4)$ with
\begin{eqnarray}
  \label{eq:currentlowesttc}
  \overline{I}^{\rm TG}\equiv -eK_0(0).
\end{eqnarray}
For a sinusoidal $\varepsilon(t)=\varepsilon + \Delta \sin(\Omega t)$, 
the explicit expression Eq.~(\ref{KGexplicit}) yields 
\begin{eqnarray}
  \label{eq:currentstatnew}
  \overline{I}^{\rm TG}
&=&-eT_c^2\sum_n J_n^2\left(\frac{{\Delta}}{\Omega}\right) \mbox{\rm Re} \left(\frac{2C_{\varepsilon+n\Omega}}
{1+\frac{\Gamma_R}{2}C_{\varepsilon+n\Omega}}\right)
\end{eqnarray}
Note that Eq.~(\ref{eq:currentstatnew}) is the Tien-Gordon formula. This can be easily demonstrated by 
expanding the non-driven stationary current,  Eq.~(\ref{currentstat}), to lowest order in $T_c$, namely
$\overline{I} =  \overline{I}_0 + O(T_c^4)$, such that, for the driven case: 
\begin{eqnarray}
  \label{eq:currentstatnew1}
  \overline{I}^{\rm TG}&\equiv&\sum_n J_n^2\left(\frac{{\Delta}}{\Omega}\right)
\left.\overline{I}_0\right|^{{\Delta} = 0}_{\varepsilon\to\varepsilon+n\Omega}.
\end{eqnarray}
To lowest order in $T_c$, the stationary current therefore is given by the
Tien-Gordon formula:
the current in the driven system is expressed by a sum over 
current contributions from side-bands $\varepsilon+n\Omega$, weighted 
with squares of Bessel functions. Note that the perturbative result $\overline{I}^{\rm TG}\equiv -eK_0(0)$, 
Eq.~(\ref{eq:currentlowesttc}), does not refer to any
specific form of the periodic function $\varepsilon(t)$; it is valid for arbitrary periodic
driving when the corresponding Fourier component  $K_0(0)$ is used.


\subsection{Non-Adiabatic Approximation}
This approximation assumes that the frequency $\Omega$ is the largest energy scale
in the problem,
\begin{eqnarray}
  \Omega\gg T_c,\varepsilon,\Gamma_R,\Gamma_L.
\end{eqnarray}
On the r.h.s. of the integral equation Eq.(\ref{eq:eomLapl3}) for $\hat{n}_{L/R}(z)$,
one then replaces the integral kernels $\hat{K}(z,t)$ and $\hat{G}(z,t)$ by their
averages over one period of the AC field,
\begin{eqnarray}
  \hat{K}(z,t) &\to& \frac{\Omega}{2\pi}\int_0^{2\pi/\Omega} dt \hat{K}(z,t) \equiv K_0(z)
\end{eqnarray}
and similarily for $\hat{G}(z,t)$. 
The Fourier coefficients
$K_n(z)$ and $G_n(z)$ with $n\ne0$ then vanish and one obtains
$\bar{I} \approx \bar{I}^{\rm fast}$, where
\begin{eqnarray}\label{currentfast}
  \bar{I}^{\rm fast} \equiv \frac{-e\Gamma_R K_0(0) }{\Gamma_R+G_0(0)+K_0(0)\left[1+\Gamma_R/\Gamma_L\right]}.
\end{eqnarray}
We observe that within lowest order of the static tunneling $T_c$, 
Eq.~(\ref{currentfast}) coincides with the Tien-Gordon expression, Eq.~(\ref{eq:currentlowesttc}), which
one obtains by setting $G_0(0)\propto T_c^2$ and $K_0(0)\propto T_c^2$ to zero in the 
denominator of Eq.~(\ref{currentfast}). In fact, for the undriven case $\Delta=0$ one can prove
\cite{BV01} that the expression for the stationary current sums up an infinite number of terms
$\propto T_c^2$, a fact that can be traced back to the 
integral equation structure of the underlying master equation.
Here, Eq.~(\ref{currentfast}) demonstrates that a similar summation effectively can be achieved
in the AC driven case.

\subsection{Higher Order Corrections to Tien-Gordon}
In order to systematically go beyond the Tien-Gordon approximation, 
Eq.~(\ref{eq:currentlowesttc}),  one has to
perform an expansion of the current in powers of $T_c^2$. This can be achieved by
{\em truncating} the infinite set of linear equations,  Eq.~(\ref{Aequation}), in order to obtain
approximations for the $n=0,\pm 1, \pm 2$-th side-band values of $\nu_n$ and $\mu_n$. The simplest way
to do this in practice is by a numerical solution of these equations as discussed below.

Barata and Wreszinski \cite{BW00} have considered higher order corrections 
to dynamical localisation in a {\em closed}, coherent two-level system, i.e., without coupling to external 
electron reservoirs or dissipation. They found that the next order in
perturbation theory given a contribution different from zero 
was the third order one, giving a contribution to a renormalisation of the 
tunnel coupling $T_c$:
\begin{eqnarray}\label{thirdorder}
  & &\delta T_c^{(3)}\equiv -\frac{2T_c^3}{\Omega^2}\times\\
& & \sum_{n_1,n_2 \in Z} 
\frac{J_{2n_1+1}\left(\frac{\Delta}{\Omega}\right)J_{2n_2+1}\left(\frac{\Delta}{\Omega}\right)
J_{-2(n_1+n_2+1)}\left(\frac{\Delta}{\Omega}\right)}
{(2n_1+1)(2n_2+1)}.\nonumber
\end{eqnarray}
We now recall our expression 
\begin{eqnarray}
  K_0(0)=\sum_n \left[ T_c J_n\left(\frac{{\Delta}}{\Omega}\right) \right] ^2
2\mbox{\rm Re} D_{\varepsilon+n\Omega}
\end{eqnarray}
(and  $G_0(0)$ correspondingly with $D_{\varepsilon+n\Omega}$ replaced by 
$E_{\varepsilon+n\Omega}$), cf.  Eq.~(\ref{KGexplicit}), 
which enter the Tien-Gordon result,  Eq.~(\ref{eq:currentlowesttc}), and the 
re-summed non-adiabatic approximation  Eq.~(\ref{currentfast}).
We use the renormalised $T_c$,  Eq.~(\ref{thirdorder}),
in order to define  a renormalised function $K_0^{(3)}(0)$,
\begin{eqnarray}
K_0^{(3)}(0)\equiv
\sum_n \left[ T_c J_n\left(\frac{{\Delta}}{\Omega}\right)  + \delta T_c^{(3)}\right] ^2
2 \mbox{\rm Re} D_{\varepsilon+n\Omega},
\end{eqnarray}
and $G_0^{(3)}(0)$ correspondingly. This yields an expression for the current, 
renormalised up to third order in $T_c$, according to
\begin{eqnarray}\label{currentthird}
  \bar{I}^{(3)} \equiv \frac{-e\Gamma_R K_0^{(3)}(0) }{\Gamma_R+G_0^{(3)}(0)+K_0^{(3)}(0)\left[1+\Gamma_R/\Gamma_L\right]}.
\end{eqnarray}

In the following, we discuss and compare our above results.

\section{Discussion}

\subsection{Comparison of Two Numerical Schemes}
In order to numerically solve the integro-differential system Eq. (\ref{eom3new}),
it is convenient to write
\begin{eqnarray}
  \exp\left( i \int_{t'}^{t}ds \varepsilon(s) \right) \equiv e^{i\varphi_t}  e^{-i\varphi_{t'}}
\end{eqnarray}
with $\varphi_t\equiv \varepsilon t -(\Delta/\Omega) \cos \Omega t$, remembering our choice
$\varepsilon(t)=\varepsilon+\Delta \sin \Omega t$.
We then introduce the real and imaginary part of
$\langle p \rangle$, use $e^{ix}=\cos x + i \sin x$,
and specify to the Ohmic dissipation case for $C(t)$,
\begin{eqnarray}\label{gammacomplex}
C(t)&=&|C(t)|e^{-i\Psi_t},\quad
  \Psi_t = 2\alpha \arctan \omega_c t\\
|C(t)| &=& [1+(\omega_c t)^2]^{-\alpha} \left| \frac{\Gamma(1+1/\beta\omega_c + it/\beta)}
{\Gamma(1+1/\beta\omega_c)}\right|^{4\alpha}.\nonumber
\end{eqnarray}
We have solved Eqs. (\ref{eom3new}) numerically
as a function of time, with the result for large times
used to obtain the stationary current as a function of $\varepsilon$.
For each value of $\varepsilon$, the time-dependent equations
have been solved up to a fixed final time $t_f$  with a subsequent
time-average over the interval $[t_f-\Delta t, t_f]$. $t_f$ has to be chosen sufficiently
large, in particular for larger values of $\alpha$. Consequently, one then also has to increase the
number of steps to achieve sufficient accuracy of the data. We have used these numerical results
to check our method for the stationary quantities as obtained from 
truncating Eq. (\ref{Aequation}) at a finite photo-sideband number, and found 
good agreement between both methods. Whereas the direct integration of the 
equations of motion is somewhat slower than the truncation method, it has the advantage that it does not 
require analytic forms of the Laplace transform for the bosonic correlation functions
$\hat{C}_\varepsilon$, Eqs. (\ref{eq:Laplacedef}) and  Eqs. (\ref{Cepsdefine}). The latter
are required for the matrix scheme Eq. (\ref{Aequation}). In Appendix \ref{bosoncorr}
we derive explicit expressions for zero temperature ($T=0$) and Ohmic dissipation.
Note that in contrast to  usual `$P(E)$' theory, we require both 
the real part $\mbox{Re}[\hat{C}_\varepsilon (0)]=\pi P(\varepsilon)$ (where $P(\varepsilon)$ is the probability for inelastic tunneling with energy transfer $\varepsilon$ \cite{Weiss}), {\em and}
the imaginary part of $\hat{C}_\varepsilon$. 

In the following, we show numerical results obtained with the truncation method. 
\begin{figure}[t]
  \includegraphics[width=1\columnwidth]{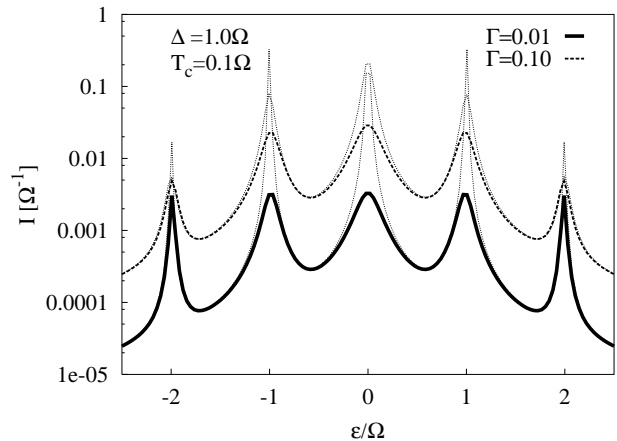}
\caption[]{\label{exact_tiengordon.eps}
Average current through double dot in Coulomb blockade regime with bias 
$\varepsilon+\Delta \sin \Omega t$. Coupling to left and right leads $\Gamma_L=\Gamma_R=\Gamma$.
Dotted lines indicate Tien-Gordon result, Eq.(\ref{eq:currentstatnew1})}
\end{figure}

\subsection{Photo-Sidebands (Coherent Case)}

\subsubsection{Comparison with Tien-Gordon approximation}
In Fig. (\ref{exact_tiengordon.eps}), we compare the exact numerical result for the
average stationary current with the Tien-Gordon expression, Eq.(\ref{eq:currentstatnew1}),
in the coherent case $\alpha=0$. One clearly recognises the symmetric photo-side peaks 
which, according to Eq.(\ref{eq:currentstatnew1}), appear at $\pm n \hbar \varepsilon$.
The Tien-Gordon approximation overestimates the current
close to these resonances, where terms of higher order in $T_c$ become
important due to the non-linearity (in $T_c$) of the exact bonding
and antibonding energies $\pm \sqrt{\varepsilon^2+4T_c^2}$ of the isolated two-level
system. This again confirms that the Tien-Gordon result is perturbative in the
tunneling $T_c$. 

\subsubsection{RWA and Bloch-Siegert shift}
\begin{figure}[t]
  \includegraphics[width=1\columnwidth]{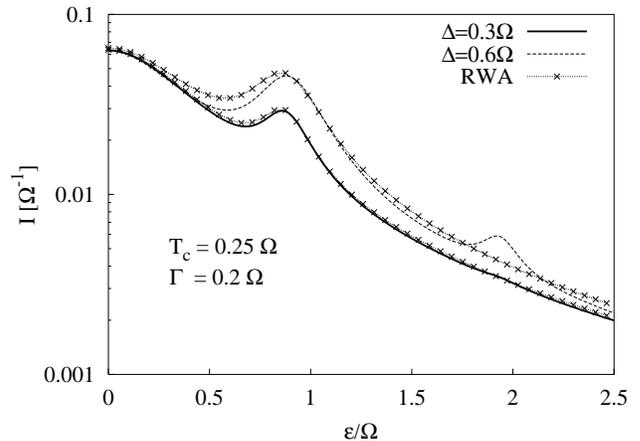}
\caption[]{\label{sidepeak.eps}
Comparison between RWA, Eq.~(\ref{SNresult}), and exact result for first current side-peak.\\
}
\end{figure}

Close to the first side-peak, Stoof and Nazarov have used a
Rotation Wave Approximation (RWA) to obtain analytical
predictions for the first current side-peak. In this
approximation, one transforms into an interaction picture where
the fast-rotating terms with angular frequency $\pm \Omega$ are
transformed away, and terms with higher rotation
frequencies (such as $\pm 2\Omega$) are neglected.
The resulting expression for the current is \cite{SN96}
\begin{eqnarray}\label{SNresult}
  I_{\rm SN} = \frac{\Delta^2\Gamma_R(a^2-4)}{c(c\Gamma_R^2+b\Delta^2)}\frac{w^2}{w^2+(\varepsilon-\varepsilon_r)^2},
\end{eqnarray}
with the resonance point $\varepsilon_R\equiv \sqrt{\Omega^2-4T_c^2}$ and 
parameters $a=\Omega/T_c$, $b\equiv\Gamma_R/\Gamma_L+2$, $c\equiv a^2+b-4$, and
the half-width $w=(a/[2\sqrt{a^2-4}])\sqrt{\Gamma_R^2+(b/c)\Delta^2}$.
We compare $I_{\rm SN}$ with the exact result in Fig.(\ref{sidepeak.eps}). 

For smaller driving amplitude $\Delta$,
the agreement is very good but becomes worse with increasing $\Delta$. The 
position of the side-peak resonance point, which is independent of
$\Delta$ in the Stoof-Nazarov approximation \ Eq.~(\ref{SNresult}), starts to
shift towards slightly larger values of the bias $\varepsilon$. 
In fact, for stronger AC driving the RWA is known to break down: in isolated
two-level systems, the first corrections to the RWA lead to the well-known
Bloch-Siegert shift \cite{Allen} of the central resonance towards larger energies, which is
consistent with the exact result in Fig. (\ref{sidepeak.eps}).


\subsection{Dynamical Localization and its Lifting}

In a quantum system driven by a periodic electric
field, a phenomenon termed Coherent Destruction of Tunneling (CDT) (also
denoted Dynamical Localization (DL)) occurs under certain parameters of 
the external field \cite{PA03,Groetal91}. The periodicity of the external field 
allows to write the solutions of the Schr\"odinger equation as:
$\psi(t) = \exp[-i \epsilon_j t] \phi_j(t)$ where $\epsilon_j$ is
called the quasi-energy, and $\phi_j(t)$ is a function with the
same period as the driving field: the Floquet state.

When two quasi-energies approach degeneracy, the time-scale for
tunneling between the states diverges, producing the phenomenon of
coherent destruction of tunneling (CDT) \cite{Groetal91}. The
time scale for localization is the inverse of the energy
separation of the quasienergies.

In the case of an isolated two level system
driven by a monochromatic, sinusoidal field 
${\varepsilon}(t)=\varepsilon + \Delta \sin(\Omega t)$, Eq.(\ref{eq:eeesin}),
CDT can be physically understood from the
renormalization of the 
coupling $T_c$ of the two levels,
\begin{equation}
T_c \rightarrow T_{c,{\rm eff}} \equiv T_c
J_0\left(\frac{\Delta}{\hbar\Omega}\right).
\label{CDT}
\end{equation}
This expression is obtained from first-order perturbation theory in the tunneling $T_c$ \cite{PA03}.
At the first zero of the Bessel function
$J_0$, namely when 
$\Delta/\hbar\Omega =2.4048...$, the {\it
effective tunnel splitting vanishes}, leading to a complete
localization of the particle in the initial state. 

In the following, we discuss how stronger tunnel
amplitudes $T_c$, the coupling to the external leads, and dissipation
modify this picture.

\begin{figure}[t]
  \includegraphics[width=1\columnwidth]{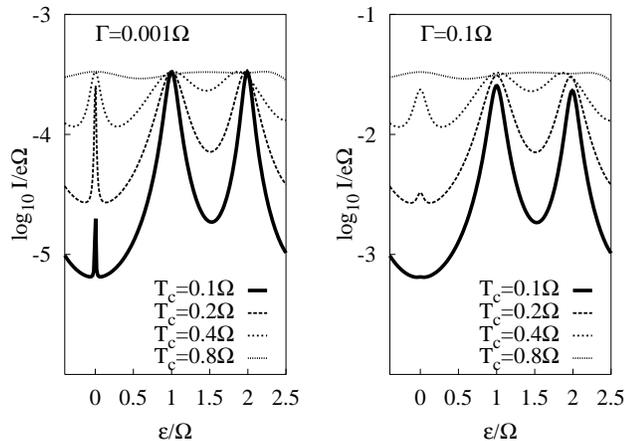}
\caption[]{\label{dlregime.eps}
Average current for AC driving amplitude $\Delta=z_0 \Omega$ ($z_0$ first zero of Bessel function $J_0$)
and various tunnel couplings $T_c$. Coupling to left and right leads $\Gamma_L=\Gamma_R=\Gamma$.
}
\end{figure}

\subsubsection{Current Suppression}
In Fig. (\ref{dlregime.eps}), we show results for the average current and $\alpha=0$ 
(no dissipation) in the dynamical localisation (DL) regime. 
Here, we define this regime by
$\Delta=z_0 \Omega$, where $z_0=2.4048...$ is the first zero of the Bessel function
$J_0$. For this specific value of the AC driving $\Delta$, to lowest order in $T_c$ the
average current is strongly suppressed for $|\varepsilon|\lesssim \Omega$
as compared with the un-driven case $\Delta=0$. 
For small $T_c$, this suppression is  
well-described by the Tien-Gordon expression
(not shown here): since at
$\Delta=z_0 \Omega$, the $n=0$ term in the sum Eq.~(\ref{eq:currentstatnew1}) is absent,
the current is dominated by the shifted (un-driven) current contributions at
bias $\varepsilon+n\Omega$ with $|n|\ge 1$, which however are 
very small due to the resonance shape of the un-driven current. \\


\subsubsection{Central Current Peak and Third-Order Result}
\begin{figure}[t]
\includegraphics[width=1\columnwidth]{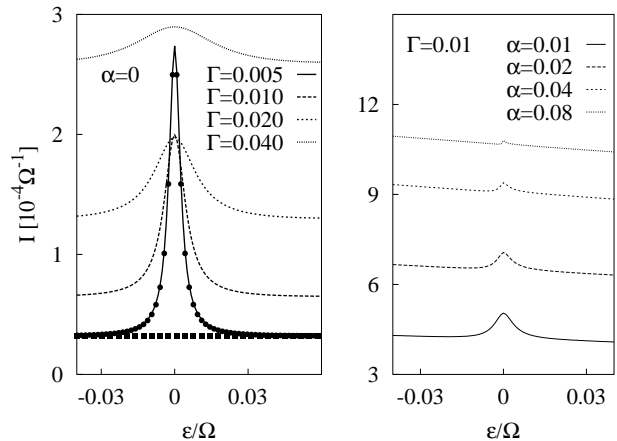}
\caption[]{\label{central.eps}Central peak of average current 
through AC driven double quantum dot. Parameters $T_c=0.1$, 
$\Delta=z_0 \Omega$ (all rates in units of $\Omega$)
LEFT: coherent case $\alpha=0$  
for different tunnel rates $\Gamma=\Gamma_L=\Gamma_R$, 
dots indicate third order results Eq.~(\ref{currentthird}),
squares indicate the Tien-Gordon result Eq.~(\ref{eq:currentstatnew1}) for the
case $\Gamma=0.005$. RIGHT:
disappearance of central peak with increasing dissipation $\alpha$.}
\end{figure}

Surprisingly, however, the coherent suppression of the current is
{\em lifted} again very close to $\varepsilon=0$, where a small and sharp peak
appears. This peak becomes broader with increasing tunnel coupling $T_c$, but
its height is suppressed for increasing reservoir coupling $\Gamma$, cf. Fig. (\ref{dlregime.eps}) right. 
This feature is analysed in
Fig. (\ref{central.eps}), where we show results for the central current peak
around $\varepsilon=0$ in the DL regime for coherent ($\alpha=0$, left) and
incoherent ($\alpha>0$, right) tunneling.
As one recognises,  the Tien-Gordon description (which is 
perturbative in the tunnel coupling $T_c$) breaks down
close to $\varepsilon=0$ where higher order terms in $T_c$ become important.
As a matter of fact, for $\varepsilon=0$ the only relevant energy scale of the 
isolated two-level systems is $T_c$ itself. In contrast, the {\em third order approximation}
Eq.~(\ref{currentthird}) reproduces very well the additional peak at $\varepsilon=0$, which indicates
the importance of higher order terms in that regime. At $\varepsilon=0$, the charge between 
the two dots is strongly de-localised in the undriven case, and this tunneling-induced quantum coherence
persists into the strongly driven regime where its signature is a `lifting' of the
DL close to $\varepsilon=0$. 

The width of the corresponding
current peak  is determined by the tunneling rate $\Gamma$. An increase of 
incoherent electron tunneling from the left lead therefore washes out the
coherent lifting of the DL.
This argument in emphasised in the right part of Fig. (\ref{central.eps}) which shows that the 
central peak in the DL regime vanishes for increasing dissipation strength $\alpha$.

\subsection{Dissipation and Average Current}
\begin{figure}[t]
\includegraphics[width=1\columnwidth]{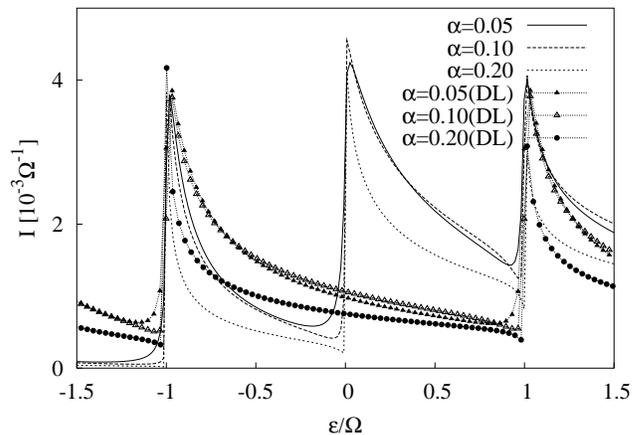}
\caption[]{\label{dissipation.eps}Average current through double dot in Coulomb blockade regime with bias 
$\varepsilon+\Delta \sin \Omega t$ for various Ohmic dissipation strengths $\alpha$ 
at zero temperature.
Driving amplitude $\Delta=\Omega$ for lines without symbols, $\Delta=z_0 \Omega$ 
($z_0$ first zero of Bessel function $J_0$) 
for lines with symbols. Tunnel coupling between dots $T_c=0.1\Omega$,
bath cutoff $\omega_c=500 \Omega$, and lead tunnel rates
$\Gamma_L=\Gamma_R=0.01\Omega$.}
\end{figure}

\subsubsection{Dissipative Photo-Sidebands}
As mentioned above, for simplicity we restrict ourselves to an Ohmic dissipative bath at
zero temperature ($T=0$) in this paper, leaving the finite temperature case
or the case of more complicated spectral functions $J(\omega)$ for future work.

For $\Delta=0$, we reproduce the analytical result  Eq.~(\ref{currentstat}) 
and the corresponding inelastic current part for $\varepsilon>0$ 
due to spontaneous boson emission \cite{Fujetal98,BK99}.
In Fig. (\ref{dissipation.eps}), we show the stationary current as a function
of bias $\varepsilon$ for various Ohmic dissipation strengths $\alpha$ 
at zero temperature and finite AC driving amplitudes $\Delta$. For
$\Delta=\Omega$, apart from the 
central resonant tunneling peak, side-bands at $\varepsilon= n\Omega$ appear which
reproduce the asymmetry of the central peak around $\varepsilon=0$. This
asymmetry  is  a clear signature of
the coupling to the dissipative environment strongly modifying the current
even at zero temperature. 

The specific form of the inelastic current depends on the boson spectral density
$J(\omega)$ \cite{BK99}.
Note that in general, there is no
monotonic dependence on the dissipation strength $\alpha$ since
the boson correlation function  $\hat{C}_\varepsilon$
appears both in the denominator and the numerator of the expression for the 
current  Eq.~(\ref{currentstat}). 

\begin{figure}[t]
\includegraphics[width=1\columnwidth]{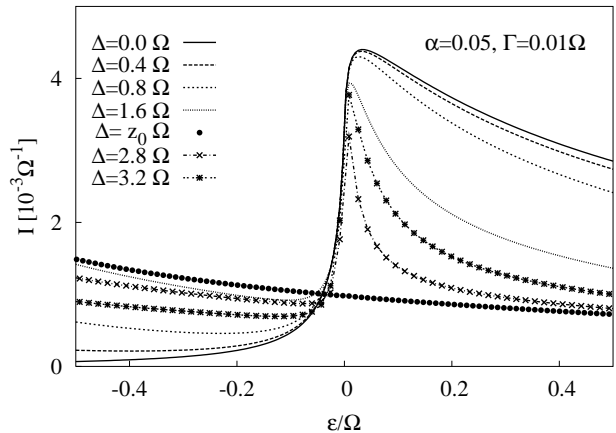}
\caption[]{\label{driving.eps}Average current through driven double dot 
for various AC driving amplitudes $\Delta$ and fixed dissipation $\alpha=0.05$, tunnel coupling
$T_c=0.1 \Omega$.}
\end{figure}

\subsubsection{Dissipation and Dynamical Localisation in the Current}
If the AC driving amplitude $\Delta$ is increased towards $z_0\Omega$ ($z_0$ is the first zero of the Bessel
function $J_0$), one expects to enter the regime of dynamical localization and a strong
suppression of the central current peak.
In the coherent case $\alpha=0$ (see above), resonant tunneling is usually strongly inhibited 
 due to coherent suppression of tunneling. 

For $\alpha>0$, however, we find that the current suppression strongly depends on the static bias
$\varepsilon$:
we find suppression for $\varepsilon>0$ and, in general, {\em larger} values 
of the current for $\varepsilon<0$ as compared to the case of smaller AC amplitudes $\Delta$.
We explain this feature in the following: 
the dependence of the average current on the driving amplitude $\Delta$ for fixed $\alpha$
is clearly visible in Fig. (\ref{driving.eps}).
A small driving amplitude $\Delta\lesssim 0.2$ nearly does not change the current at all. However,  
the originally strongly asymmetric current curve becomes flattened out when
$\Delta$ is tuned to larger values up to the dynamical localization value $\Delta=z_0\Omega$.
There, the AC field nearly completely destroys the strong asymmetry 
between the spontaneous emission ($\varepsilon>0$) and the absorption side ($\varepsilon<0$)
of the current. The central $n=0$ photo-band is completely suppressed and the 
dominant contribution to the current comes from the $n=\pm 1$ bands. 
For $\varepsilon<0$, the current for $\Omega>|\varepsilon|$ is due to 
photo-excitation of the electron
into the first upper photo-sidebands and subsequent spontaneous emission of bosons 
of energy $E_1\equiv \Omega-|\varepsilon|$ to the bath.
In contrast, for $\Omega>\varepsilon>0$, photon emission blocks the current
because  at $T=0$ there is no absorption of bosons from the bath. The remaining 
photon absorption channel then leads to boson emission at an energy 
$E_2\equiv\Omega+\varepsilon$, which is larger as compared to the case
for $\varepsilon<0$, namely $E_2>E_1$, and therefore has
a smaller probability
$P(E)\propto E^{2\alpha-1}e^{-E/\omega_c}$, cf. Eq.(\ref{PE}), leading to a smaller current.
A similar argument can be used to explain why the current {\it increases} as one reduces
$\varepsilon$, say from $\varepsilon/\Omega=0$ to $\varepsilon/\Omega=-0.5$.
In particular, the strongest effect of the dissipative bath occur near one-photon 
resonance conditions, i.e. when $\varepsilon/\Omega \approx\pm 1$, where
the current is regulated by the spectral function
of the bosonic bath at very low frequencies, either by absorption ($\varepsilon<0$) 
or emission ($\varepsilon>0$) of a photon. These processes appear in the current as
nonanalytic cusps reflecting the power law behavior of $P(E)$. This has to be compared 
with the Lorentzian shape of the photo-sidebands in the absence of dissipation (Fig.~1).
If one tunes to even larger values of $\Delta>z_0\Omega$, 
the central $n=0$ photo-band re-appears and the original strong asymmetry of the current curve
is restored.

\section{Conclusion and Outlook}
Our results suggest that the combination of AC fields and dissipation in double
quantum dots leads to a rich variety of non-trivial effects. In particular, we have 
shown that a time-dependent monochromatic field drastically modifies the 
dissipative inelastic stationary current, in particular for stronger AC driving in
the dynamical localisation regime. Corrections to the Tien-Gordon formula
appear at larger tunnel coupling between the dots and become extremely
important near zero bias in the DL regime, also in the non dissipative
case.

The method presented in this work has the benefit of
accounting for an arbitrary dissipative
environment via the correlation function $\hat{C}_\varepsilon$.
In the generic case, explicit analytical forms for this function
are difficult to obtain and it might be easier to integrate the 
original  equations of motion directly. Alternatively, one can numerically
evaluate $\hat{C}_\varepsilon$ and use it as an input into our Floquet-like formalism.
We also mention that the entire approach is based on the decoupling 
of the bosonic degrees of freedom  in the polaron transformed master equation. One is therefore 
always restricted to  the range of validity of the NIBA (non-interacting blip approximation) 
of the original spin-boson problem \cite{Weiss,GH98}. 
Discussing larger temperatures $T$ should thus lead to 
more reliable results as compared to the `test-models' $C_\varepsilon$ 
which were discussed here for $T=0$.

A future extension of our approach
should therefore be the derivation of a systematic perturbation theory in the 
electron-boson coupling, starting from the bonding-antibonding basis of the 
double dots. In a calculation for
an undriven double quantum dot, such an approach has been successfully used recently to extract
dephasing and relaxation times from the frequency dependent noise spectrum \cite{AB03}.

Even for the coherent case $\alpha=0$, our results have shown that there are
non-trivial effects due to the combined quantum coherence inherent in the 
double dot, and the coherence induced by the external driving field. In particular, we
found systematic corrections to standard approximation such as the Tien-Gordon formula 
or the rotating wave approximation. The constituing quantities $K_m$ and $G_m$ 
of our theory,  cf. Eq.~(\ref{currentstationary},\ref{KGexplicit}),
describe dissipative tunneling $\propto T_c^2$ of one additional quasi-particle between the two dots
under the influence of the AC field, which again indicates that our approach is
essentially perturbative in $T_c$, although to infinite order and exact for $\alpha=0$.
We showed that partial re-summations beyond the Tien-Gordon result
are justified in a non-adiabatic, high-frequency approximation,
but for the general case one has to rely on a systematic evaluation of Eq.~(\ref{currentstationary}).

This work was supported by DFG BR 1528,
EPSRC GR/R44690, the UK Quantum Circuits Network, the Nuffield foundation (T.B.),  by the  
MCYT of Spain through the "Ram\'on y Cajal" program (R.A.) and 
project MAT2002-02465 (G.P. and R.A.). We acknowledge as well The EU Human
Potential Programme under contract HPRN-CT-2000-00144.

\begin{appendix}
\section{Dot Occupancies in Laplace Space}
Here, we derive  Eq.~(\ref{eq:eomLapl3}) for the occupancies $\langle n_{L/R} \rangle$.
We define
\begin{eqnarray}
  q(t)&\equiv& \langle p \rangle _t e^{-i\int_{0}^{t}ds\,\tilde{\varepsilon}(s)},\quad
  q^{\dagger}(t)\equiv \langle p^{\dagger} \rangle _t e^{+i\int_{0}^{t}ds\,\tilde{\varepsilon}(s)}\nonumber\\
\end{eqnarray}
This is inserted into the equations of motion in the time domain,  Eq.~(\ref{eom3new}), 
which upon Laplace transformation become
\begin{widetext}
\begin{eqnarray}
  \label{eq:eomLapl1}
  z\hat{n}_L(z)-\langle n_L \rangle_0 &=& -i \int_{0}^{\infty}dt e^{-zt} \left\{
T_c(t)q(t)- T_c^*(t) q^{\dagger}(t) \right\} 
+\Gamma_L\left[ \frac{1}{z} - \hat{n}_L(z)- \hat{n}_R(z) \right]
\nonumber\\
z\hat{n}_R(z)-\langle n_R \rangle_0 &=& i \int_{0}^{\infty}dt e^{-zt} \left\{
T_c(t)q(t)- T_c^*(t) q^{\dagger}(t) \right\} 
-{\Gamma}_R \hat{n}_R(z) 
\nonumber\\
\hat{q}(z)&=&-\frac{\Gamma_0}{2}\hat{q}(z)\hat{C}_{\varepsilon}(z)
-i\left[\int_{0}^{\infty}dt'e^{-zt'}T_c^*(t')\left[\langle n_L\rangle_{t'}\hat{C}_{\varepsilon}(z)
-\langle n_R\rangle_{t'}\hat{C}^*_{-\varepsilon}(z)\right]\right]\nonumber\\
\hat{q}^{\dagger}(z)&=&-\frac{\Gamma_0}{2}\hat{q}^{\dagger}(z)\hat{C}_{\varepsilon}^*(z)
+i\left[\int_{0}^{\infty}dt'e^{-zt'}T_c(t')\left[\langle n_L\rangle_{t'}\hat{C}^*_{\varepsilon}(z)
-\langle n_R\rangle_{t'}\hat{C}_{-\varepsilon}(z)\right]\right],
\end{eqnarray}
\end{widetext}
where we used the convolution theorem in the equations for $\hat{q}(z)$ and $\hat{q}^{\dagger}(z)$
and the definitions Eq.~(\ref{eq:Laplacedef}).
Using the definitions for the propagators $D$ and $E$,  Eq.~(\ref{eq:Ddef})
we obtain upon solving for $\hat{q}^{(\dagger)}(z)$ and Laplace back-transforming,
\begin{widetext}
\begin{eqnarray}
  \label{eq:qeqn1}
  q(t)&=&-i\int_{0}^{t}dt'T_c^*(t')\left[\langle n_L\rangle_{t'}{D}_{\varepsilon}(t-t')
-\langle n_R\rangle_{t'}{E}_{\varepsilon}(t-t')\right]\nonumber\\
  q^{\dagger}(t)&=&i\int_{0}^{t}dt'T_c(t')\left[\langle n_L\rangle_{t'}{D}^*_{\varepsilon}(t-t')
-\langle n_R\rangle_{t'}{E}^*_{\varepsilon}(t-t')\right],
\end{eqnarray}
\end{widetext}
involving the propagators in the time-domain.
Insertion into Eq. (\ref{eq:eomLapl1}) yields
\begin{widetext}
\begin{eqnarray}
  \label{eq:eomLapl2}
  z\hat{n}_L(z)-\langle n_L \rangle_0 &=& - \int_{0}^{\infty}dt e^{-zt}
\int_{0}^{t}dt' 
\langle n_L\rangle_{t'}\left[T_c(t)T_c^*(t'){D}_{\varepsilon}(t-t')+
T_c^*(t)T_c(t'){D}^*_{\varepsilon}(t-t')\right]\\
&+&
\int_{0}^{\infty}dt e^{-zt}
\int_{0}^{t}dt' 
\langle n_R\rangle_{t'}\left[T_c(t)T_c^*(t'){E}_{\varepsilon}(t-t')+
T_c^*(t)T_c(t'){E}^*_{\varepsilon}(t-t')\right]
+\Gamma_L\left( \frac{1}{z} - \hat{n}_L(z)- \hat{n}_R(z) \right).\nonumber
\end{eqnarray}
\end{widetext}
At this point, it is useful to use a relation for a 
generalized convolution of a function $K(t,t')$ and $f(t')$,
\begin{eqnarray}\label{convolution}
& &\int_{0}^{\infty}dt e^{-zt}\int_{0}^{t}dt' K(t,t')f(t')=\nonumber\\
&=&\int_{0}^{\infty}dt e^{-zt}f(t)\int_{0}^{\infty}dt'e^{-zt'}K(t+t',t)
\end{eqnarray}
which can be easily proven by substitutions.
Note that the usual Laplace convolution theorem is recovered from 
Eq.~(\ref{convolution}) if $K(t,t')=K(t-t')$ is only a function of the difference of 
its two arguments.
Eq.~(\ref{eq:eomLapl2}) and a similar equation for $\hat{n}_R(z)$ then lead to 
Eq.~(\ref{eq:eomLapl3}).


\section{Calculation of the Boson Correlation Function}
\label{bosoncorr}

Explicit expressions for the bosonic correlation functions
$\hat{C}_\varepsilon$, Eqs. (\ref{eq:Laplacedef}) and  Eqs. (\ref{Cepsdefine}), which can be
obtained in the zero temperature ($T=0$) case for Ohmic dissipation.
In this case,
\begin{eqnarray}
  J(\omega)&=&2\alpha \omega \exp(-\omega{}/\omega{}_c)\nonumber\\
C(t)&=&(1+i\omega_c t)^{-2\alpha},\quad g\equiv 2\alpha.
\end{eqnarray}
We have
\begin{eqnarray}
  \hat{C}(z)&\equiv&\int_{0}^{\infty}dt e^{-zt}(1+i\omega_c t)^{-2\alpha}\\
&=&(i\omega_c)^{-2\alpha}z^{2\alpha-1}e^{-iz/\omega_c}\Gamma(1-2\alpha,-iz/\omega_c),\nonumber
\end{eqnarray}
where we used Gradstein-Ryshik 3.382.4 and $\Gamma$ denotes the incomplete Gamma function.
We set $\omega_c=1$ for a moment to simplify notations and obtain
\begin{eqnarray}
  \hat{C}(-i\varepsilon)&=&-i\left(-\varepsilon\right)^{2\alpha-1}
e^{-\varepsilon}
\Gamma(1-2\alpha,-\varepsilon)
\end{eqnarray}
Note that $\varepsilon$ must have a small positive imaginary part here ($\mbox{Re} z>0$ in the definition of
the Laplace transformation): the incomplete Gamma function $\Gamma(1-2\alpha,z)$ has a branch point at 
$z=0$. However, we can use the series expansion
\begin{eqnarray}
  \Gamma(1-2\alpha,x)&=&\Gamma(1-2\alpha)-\sum_{n=0}^{\infty}\frac{(-1)^nx^{1-2\alpha+n}}{n!(1-2\alpha+n)}
\nonumber\\
1-2\alpha&\ne& 0,-1,-2,...
\end{eqnarray}
to obtain
\begin{eqnarray}
   \hat{C}(-i\varepsilon)&=&-i(-\varepsilon)^{2\alpha-1}e^{-\varepsilon}\Gamma(1-2\alpha)\\
&+&ie^{-\varepsilon}\sum_{n=0}^{\infty}\frac{\varepsilon^{n}}{n!(1-2\alpha+n)},\quad 2\alpha\ne 1,2,3,...
\nonumber
\end{eqnarray}
The second term is an analytic function  of $\varepsilon$.

Now write
\begin{eqnarray}
  -i(-\varepsilon)^{2\alpha-1}&=&\left\{
  \begin{array}{cc}
-i|\varepsilon|^{2\alpha-1},&\varepsilon<0\\
\varepsilon^{2\alpha-1}e^{-\pi i(1/2+2\alpha-1)}      &\varepsilon>0.
  \end{array}\right.\nonumber\\
  &=&\left\{\begin{array}{cc}
-i|\varepsilon|^{2\alpha-1},&\varepsilon<0\\
\varepsilon^{2\alpha-1}
\left( \sin 2\pi\alpha +i \cos 2\pi \alpha\right)
 &\varepsilon>0.
  \end{array}\right.\nonumber\\
\end{eqnarray}
Recall the reflection formula for the Gamma function,
\begin{eqnarray}
  \Gamma(1-z)=\frac{\pi}{\Gamma(z)\sin \pi z}.
\end{eqnarray}
This yields
\begin{eqnarray}
\varepsilon>0:\quad   \hat{C}(-i\varepsilon)&=&
\frac{\pi}{\Gamma(2\alpha)}\varepsilon^{2\alpha-1}e^{-\varepsilon} \\
&+& i\left[ \frac{\pi}{\Gamma(2\alpha)}\varepsilon^{2\alpha-1}e^{-\varepsilon} \cot 2\pi \alpha\right.\nonumber\\
&+&\left.
e^{-\varepsilon}\sum_{n=0}^{\infty}\frac{\varepsilon^{n}}{n!(1-2\alpha+n)}\right].\nonumber\\
\varepsilon<0:\quad   \hat{C}(-i\varepsilon)&=&
ie^{-\varepsilon}\left[-\frac{\pi}{\Gamma(2\alpha)\sin 2\pi\alpha}
|\varepsilon|^{2\alpha-1}\right.\nonumber\\
&+&\left.
\sum_{n=0}^{\infty}\frac{\varepsilon^{n}}{n!(1-2\alpha+n)}\right].\nonumber
\end{eqnarray}
From this, we can read off the real and the imaginary part of $\hat{C}(-i\varepsilon)$.
The real part is
\begin{eqnarray}\label{PE}
  \mbox{Re} \hat{C}(-i\varepsilon)&\equiv& \pi P(\varepsilon)
=\frac{\pi}{\Gamma(2\alpha)}\varepsilon^{2\alpha-1}e^{-\varepsilon} \theta(\varepsilon).
\end{eqnarray}
The imaginary part is
\begin{eqnarray}
   \mbox{Im} \hat{C}(-i\varepsilon)&\equiv&
e^{-\varepsilon}\left[\sum_{n=0}^{\infty}\frac{\varepsilon^{n}}{n!(1-2\alpha+n)}\right.\\
&+&\left.\frac{\pi|\varepsilon|^{2\alpha-1}}{\Gamma(2\alpha)\sin 2\pi\alpha}\cdot
\left\{
  \begin{array}{ll}
-1,&\varepsilon<0\\
\cos 2\pi \alpha ,&\varepsilon>0
  \end{array}\right\}
\right].\nonumber
\end{eqnarray}

\end{appendix}


\end{document}